\documentclass{jnst}
\usepackage[tablesfirst,notablist]{endfloat}
\usepackage{times}
\usepackage{txfonts}
\usepackage{graphicx}
\usepackage{footmisc}
\newcommand{\ie}{\textit{i.e.,}}
\newcommand{\eg}{\textit{e.g.,}}
\newcommand{\etal}{\textit{et al}}

\title{
EXFOR-based simultaneous evaluation for neutron-induced fission cross section of plutonium-242
}
\author{
Riko Okuyama$^{1,2}$
\thanks{Email: okuyama.riko.g0@elms.hokudai.ac.jp},
Naohiko Otuka$^{1,3}$ 
\thanks{Email: n.otsuka@iaea.org},
Go Chiba$^{2}$
\thanks{Email: go\_chiba@eng.hokudai.ac.jp}
and
Osamu Iwamoto$^{4}$ 
\thanks{Email: iwamoto.osamu@jaea.go.jp}
}
\inst{
$^{1}$Nuclear Data Section, 
      Division of Physical and Chemical Sciences,
      Department of Nuclear Sciences and Applications,
      International Atomic Energy Agency,
      A-1400 Wien, Austria;\\
$^{2}$Graduate School of Engineering,
      Hokkaido University,
      Kita-ku, Sapporo 060-0813, Japan;\\
$^{3}$Nishina Center for Accelerator-Based Science,
      RIKEN,
      Wako 351-0198, Japan;\\
$^{4}$Nuclear Data Center,
      Nuclear Science and Engineering Center,
      Japan Atomic Energy Agency,
      Tokai-mura, Naka-gun, Ibaraki 319-1195, Japan
}
\abst{
The $^{242}$Pu neutron-induced fission cross section was evaluated from 100~keV to 200~MeV.
The experimental $^{242}$Pu and $^{235}$U fission cross sections and their ratios in the EXFOR library were reviewed and analysed by the least-squares method.
Additional simultaneous evaluation was performed by including the experimental database of the $^{233,238}$U and $^{239,240,241}$Pu fission cross sections and their ratios developed for JENDL-5 evaluation.
The $^{242}$Pu fission cross sections from our evaluation and JENDL-5 evaluation are close to each other below 1~MeV while systematically differ from each other above 10~MeV.
The cross section from our evaluation is systematically lower than the JENDL-4.0 cross section in the prompt fission neutron spectrum peak region ($\sim$5\% lower around 1~MeV). 
The newly evaluated $^{242}$Pu fission cross section was verified against the cross section measured in the $^{252}$Cf spontaneous fission neutron field and criticalities of small-sized LANL fast systems,
and demonstrated better performance than the JENDL-4.0 cross section on the same level with the JENDL-5 cross section.
}
\kword{
plutonium-242,
fission,
simultaneous evaluation,
JENDL,
EXFOR
}
\begin{document}
\maketitle
\newpage
\section{Introduction}
Development of innovative nuclear systems for spent reactor fuel management such as the accelerator driven system requires high quality cross sections of plutonium and minor actinides formed by consecutive neutron captures and decays of the nuclides in the reactor fuels~\cite{Salvatores2005Nuclear}.
Target accuracies of the cross sections based on sensitivity analysis and cross section covariance are established~(\eg \cite{Alberti2006Nuclear,Salvatores2008Uncertainty}),
and efforts to improve the accuracies have been made by cross section measurements (\eg~\cite{Igashira2014Nuclear}) and evaluations~(\eg~\cite{Iwamoto2009JENDL}) as well as integral measurements (\eg~\cite{Mukaiyama1986Actinide}) and benchmark calculations (\eg~\cite{Chiba2011JENDL-4.0,Tada2023JENDL-5}).

Plutonium-242 has long half-life (3.73$\times 10^5$ year~\cite{Martin2022Nuclear}) and is accumulated during operation of thermal reactors.
Several more neutron captures and decays are required to convert $^{242}$Pu to a fissile.
However, the thermal cross section of $^{242}$Pu is rather small (18.9$\pm$0.7~b for capture, 0.16$\pm$0.04~b for fission~\cite{Mughabghab2018Atlas}),
and fast reactor systems are more efficient for transmutation of this nuclide~\cite{Tovesson2009Neutron}.
Sensitivity and uncertainty studies show reduction of the uncertainty is required for the $^{242}$Pu fission cross section in the fast neutron region.
For example, a comprehensive analysis~\cite{Salvatores2008Uncertainty} sets the target accuracy of the fission cross section to 3 -- 5\% in two energy groups (498~keV -- 1.35~MeV and 1.35~MeV -- 2.23~MeV),
where the current accuracy is $\sim$20\% and $\sim$10\% according to ``ANL covariance" or ``BOLNA covariance", respectively.
This requirement stimulated several new measurements of the $^{242}$Pu fission cross section in the fast neutron region at PTB~\cite{Belloni2022Neutron}, HZDR~\cite{Koegler2019Fast}, CEA~\cite{Marini2017242Pu}, NPL~\cite{Matei2017Absolute}, EC-JRC~\cite{Salvador-Castineira2015bNeutron} and LANL~\cite{Tovesson2009Neutron}.

It has been known that calculations with the JENDL-4.0 library~\cite{Shibata2011Japanese} overestimate the criticalities of $^{242}$Pu-loaded small-sized LANL fast systems ~\cite{Chiba2011JENDL-4.0} and $^{242}$Pu/$^{239}$Pu fission rate ratios of the FCA-IX experiment~\cite{Fukushima2017Analyses}.
In order to improve the situation,
the $^{242}$Pu fission cross section was reevaluated for the JENDL-5 library by least-squares analysis of experimental datasets including those from the above-mentioned new experiments except for the latest ones~\cite{Belloni2022Neutron,Koegler2019Fast},
and the reevaluated cross section became lower than the JENDL-4.0 cross section in the fission neutron spectrum peak region~\cite{Iwamoto2023Japanese}.

The JENDL-5 $^{242}$Pu fission cross section evaluation (2019) adopted the absolute cross sections published between 2000 and 2018 as the experimental data input.
Some of these experimental datasets were measured as the $^{242}$Pu/$^{235}$U fission cross section ratios and normalised by the experimentalists to the absolute $^{242}$Pu fission cross sections by multiplying a reference (standard) $^{235}$U fission cross section (\eg~\cite{Carlson2018Evaluation,Carlson2009International,IAEA2007International}).
The experimental data input of the JENDL-5 evaluation therefore depends on choice of the $^{235}$U reference fission cross section.
The simultaneous evaluation approach (\eg \cite{Otuka2022aEXFOR}) uses experimental cross section ratios without such conversion,
and this technique would be a more adequate for evaluation of the $^{242}$Pu fission cross section.
Also, some experimental datasets published before 2000 are well documented but discarded in JENDL-5 evaluation,
and it would be worthwhile to perform reevaluation considering these datasets as well as the experimental datasets published after JENDL-5 evaluation~\cite{Belloni2022Neutron,Koegler2019Fast}\footnote
{K\"ogler \etal's experiment~\cite{Koegler2019Fast} is not used in JENDL-5 evaluation though the dataset is plotted in Fig.~22 of the JENDL-5 evaluation full summary~\cite{Iwamoto2023Japanese}.}.

The purpose of this work is to evaluate the $^{242}$Pu fission cross section for fast neutrons by the simultaneous evaluation approach including some experiments not take into account in JENDL-5 evaluation and to check if the newly evaluated cross section still shows better performance than the JENDL-4.0 $^{242}$Pu fission cross section.

\section{Method}
We performed the present evaluation with the least-squares analysis code SOK~\cite{Kawano2000Simultaneous,Kawano2000Evaluation} which updated a prior estimate by including experimental datasets one-by-one.
We adopted the JENDL-5 cross section as the prior estimate with 50\% uncertainty without correlation.
The procedure of the JENDL-5 simultaneous evaluation~\cite{Otuka2022aEXFOR} was applied to the present evaluation,
and we provide below only brief description of general procedures and items particular for the present evaluation.

We first performed simultaneous evaluation with an experimental database collecting the $^{242}$Pu and $^{235}$U fission cross sections and their ratios (``two-nuclide evaluation") and then performed more extensive evaluation with an experimental database extended to $^{233,238}$U and $^{239,240,241}$Pu fissions (``seven-nuclide evaluation").
We set the incident energy range of the present evaluation to 100~keV to 200~MeV.
All $^{242}$Pu and $^{242}$Pu/$^{235}$U experimental datasets published no earlier than 1970 were extracted from the EXFOR library~\cite{Otuka2014Towards} and were included in our experimental database but excluding
\begin{itemize}
\item the data measured relative to a reference fission cross section which ratio is not available (\eg~\cite{Salvador-Castineira2015bNeutron,Tovesson2009Neutron,Weigmann1985Neutron})
\item the data measured with neutrons from nuclear explosions (\eg~\cite{Bergen1971Neutron,Auchampaugh1971Neutron})
\item the data compiled without partial uncertainty information (\eg~\cite{Khan198014.8})
\item the data not available from the authors and read from figure images (\eg~\cite{Shpak1990Angular,Fursov1973Anomaly})
\end{itemize}
Canc\'{e} \etal. report two $^{242}$Pu absolute cross section datasets derived with the neutron flux determined by a proton recoil telescope and BF$_3$ counter.
We combined these two datasets (EXFOR 21821.004 and 005) into one dataset (EXFOR 51013.002) to avoid use of the $^{242}$Pu fission counts from this experiment twice.
Except for this case, all numbers of the $^{242}$Pu and $^{242}$Pu/$^{235}$U experimental datasets were directly taken from the EXFOR library and converted to the SOK input files by using the SOX code~\cite{Otuka2022bEXFOR}.
For construction of the energy-energy correlation coefficients of each experimental dataset,
we assumed the uncertainty due to counting statistics is uncorrelated while the other uncertainties are fully correlated.
The experimental $^{242}$Pu and $^{242}$Pu/$^{235}$U datasets adopted in the present evaluation are listed in Table~\ref{tab:explist1}.

\begin{table*}
\caption{
$^{242}$Pu fission cross sections and their ratio to $^{235}$U fission cross sections included in the experimental database for the present evaluation.
Meadows's 1978 dataset~\cite{Meadows1978Fission} was treated as a shape dataset.
``Ver.", ``Lab." and ``Pts." give the date (N2) of the SUBENT record in EXFOR, EXFOR/CINDA abbreviation~\cite{Schwerer2014EXFOR} of the institute where the experiment was performed, and number of data points, respectively.
}
\label{tab:explist1}
\centering
\begin{tabular}{lllllrllll}
\hline
EXFOR \#   & Ver.     & First author           & Year & Lab.     & Pts.&  \multicolumn{2}{c}{Energy range (eV)} & Ref.  \\
\hline              
\textit{$^{242}$Pu}\\
23653.003.1& 20221126 & F.Belloni              & 2022 & 2GERPTB &   2 & 2.5E+06 & 1.5E+07 &\cite{Belloni2022Neutron}\\
23334.002  & 20170410 & C.Matei                & 2017 & 2UK NPL &   5 & 1.0E+06 & 2.5E+06 &\cite{Matei2017Absolute}\\
23391.003  & 20191113 & P.Marini               & 2017 & 2FR BRC &   4 & 1.0E+06 & 1.9E+06 &\cite{Marini2017242Pu}\\
31711.004  & 20110511 & K.Gul                  & 1986 & 3PAKNIL &   1 & 1.5E+07 & 1.5E+07 &\cite{Gul1986Measurements}\\
51013.002  & 20230104 & M.Cance                & 1982 & 2FR BRC &   2 & 2.5E+06 & 2.5E+06 &\cite{Cance1983Mesures}\\
\textit{$^{242}$Pu/$^{235}$U }\\
23469.002.1& 20221219 & T.K\"{o}gler           & 2019 & 2GERZFK & 238 & 5.0E+05 & 1.0E+07 &\cite{Koegler2019Fast}\\
13801.004  & 20170724 & P.Staples              & 1998 & 1USALAS & 210 & 5.1E+05 & 2.5E+08 &\cite{Staples1998Neutron}\\
22211.003  & 20200925 & T.Iwasaki              & 1990 & 2JPNTOH &  17 & 6.0E+05 & 6.8E+06 &\cite{Iwasaki1990Measurement}\\
13134.010.1& 20170724 & J.W.Meadows            & 1988 & 1USAANL &   1 & 1.5E+07 & 1.5E+07 &\cite{Meadows1988Fission}\\
22282.008.1& 20130924 & F.Manabe               & 1988 & 2JPNTOH &   3 & 1.4E+07 & 1.5E+07 &\cite{Manabe1988Measurements}\\
40509.003  & 20221221 & V.M.Kupriyanov         & 1979 & 4RUSFEI &  73 & 1.3E+05 & 7.4E+06 &\cite{Kupriyanov1979Measurement}\\
40509.005.2& 20210920 & V.M.Kupriyanov         & 1979 & 4RUSFEI &   5 & 9.8E+05 & 3.0E+06 &\cite{Kupriyanov1979Measurement}\\
10597.003  & 20221221 & J.W.Behrens            & 1978 & 1USALRL & 140 & 9.7E+04 & 3.3E+07 &\cite{Behrens1978Measurements}\\
10734.003.1& 20221223 & J.W.Meadows            & 1978 & 1USAANL &  48 & 4.0E+05 & 9.9E+06 &\cite{Meadows1978Fission}\\
\hline
\end{tabular}
\end{table*}

The energy ranges of the $^{233,235,238}$U and $^{239,240,241}$Pu cross sections considered in the present evaluation are same as those of the JENDL-5 simultaneous evaluation, namely 10~keV (fissile) or 100~keV (non-fissile) to 200~MeV,
and their experimental database constructed for the JENDL-5 evaluation~\cite{Otuka2022bEXFOR} was adopted after exclusion of one $^{235}$U dataset (See Ref.~\cite{Otuka2023Simultaneous} for the reason) and addition of a few datasets as listed in Table~\ref{tab:explist2}.

\begin{table*}
\caption{
Update of the experimental database for $^{233,235,238}$U and $^{239,240,241}$Pu fission cross sections and their ratios from the database for JENDL-5 evaluation~\cite{Otuka2022bEXFOR}.
Arlt \etal.'s ratio was deleted  while the other datasets were added.
``Ver.", ``Lab." and ``Pts." give the date (N2) of the SUBENT record in EXFOR, EXFOR/CINDA abbreviation~\cite{Schwerer2014EXFOR} of the institute where the experiment was performed, and number of data points, respectively.
}
\label{tab:explist2}
\centering
\begin{tabular}{lllllrllll}
\hline
EXFOR \#   & Ver.     & First author           & Year & Lab.     & Pts.&  \multicolumn{2}{c}{Energy range (eV)} & Ref.  \\
\hline
\textit{$^{235}$U}\\
31833.002  & 20201103 & R.Arlt                 & 1980 & 2GERZFK  &    1& 8.2E+06 & 8.2E+06 &\cite{Arlt1981Absolute}\\
23294.004  & 20211216 & I.Duran                & 2019 & 2ZZZCER  &    8& 7.5E+03 & 2.7E+04 &\cite{Duran2019High}\\
\textit{$^{240}$Pu}\\
51014.002  & 20221222 & F.Belloni              & 2022 & 2GERPTB  &    4& 2.5E+06 & 1.5E+07 &\cite{Belloni2022Neutron}\\
\textit{$^{238}$U/$^{235}$U}\\
32886.003.1& 20230206 & Z.Ren                  & 2023 & 3CPRINP  &  135& 5.1E+05 & 1.8E+08 &\cite{Ren2023Measurement}\\
\textit{$^{239}$Pu/$^{235}$U}\\
14721.002  & 20211220 & L.Snyder               & 2021 & 1USALAS  &  119& 1.1E+05 & 9.7E+07 &\cite{Snyder2021Measurement}\\
\hline
\end{tabular}
\end{table*}

\section{Results}
The number of the experimental data points used in the present evaluation is 8395 (including 1169 data points of $^{242}$Pu and $^{235}$U)
and the number of the fitting parameters was 589 (including 158 parameters of $^{242}$Pu and $^{235}$U).
In addition to the fitting parameters representing the evaluated cross sections on the energy nodes,
there are three parameters normalising the shape datasets of $^{242}$Pu/$^{235}$U (EXFOR 10734.003.1) and $^{238}$U (14529.002, 40483.002)\footnote{
See Ref.~\cite{Otuka2022aEXFOR} for treatment of EXFOR 14529.002 and 40483.002 in our evaluation.}.
With these parameters and experimental data points in the whole energy range for fitting (7 or 70~keV to 250~MeV),
the reduced chi-square (chi-square divided by the degree-of-freedom) is 3.85 for the seven-nuclide evaluation and 3.30 for the two-nuclide evaluation.

\subsection{Point-wise cross sections}
\begin{figure}
\centering\includegraphics[clip,angle=-90,width=\linewidth]{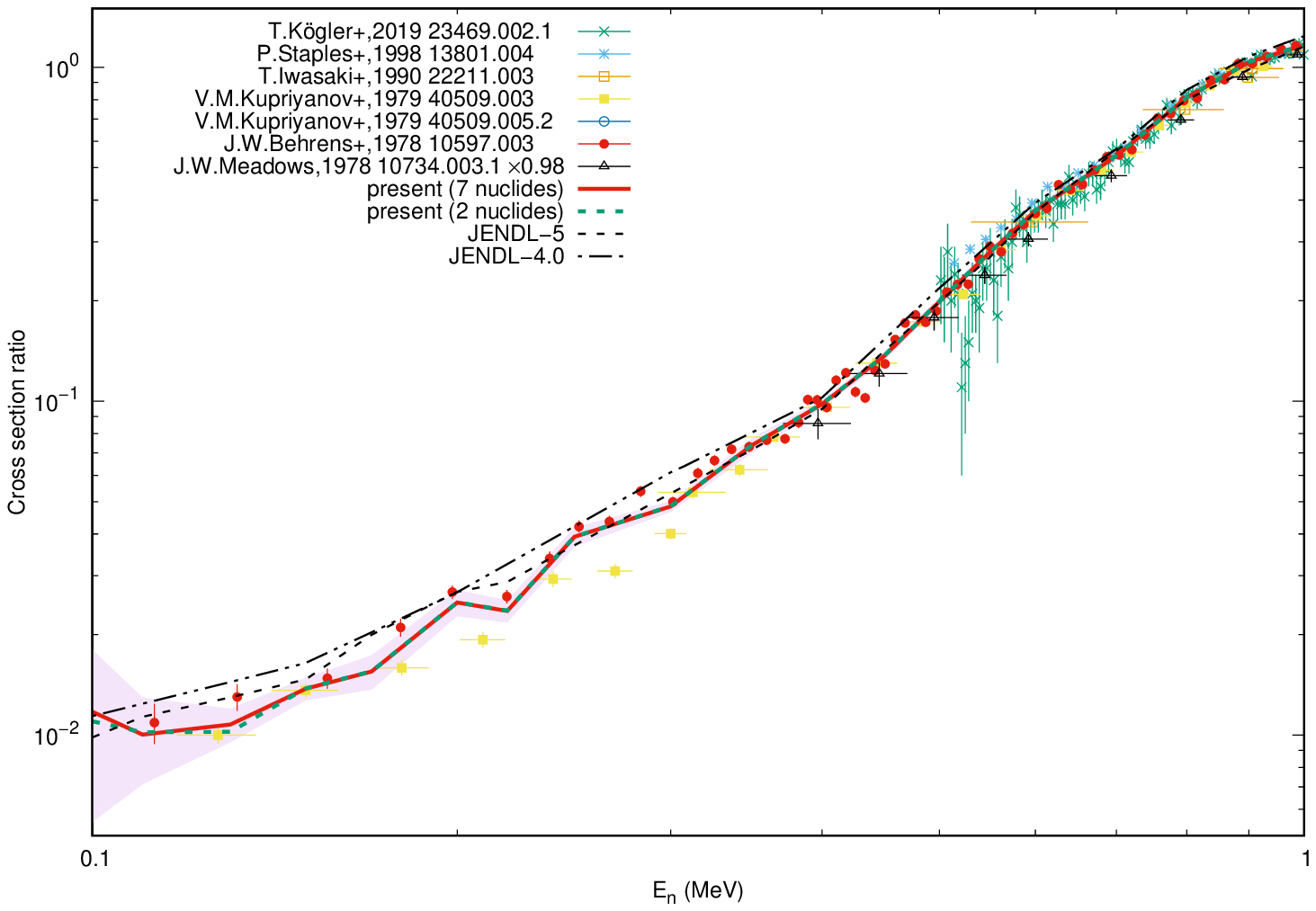}
\caption{$^{242}$Pu/$^{235}$U fission cross section ratios below 1~MeV from evaluations along with the experimental ones used in the present evaluation\cite{
Koegler2019Fast,
Staples1998Neutron,
Iwasaki1990Measurement,
Kupriyanov1979Measurement,
Behrens1978Measurements,
Meadows1978Fission
}.
}
\label{fig:Pu242U235-low}
\end{figure}

\begin{figure}
\centering\includegraphics[clip,angle=-90,width=\linewidth]{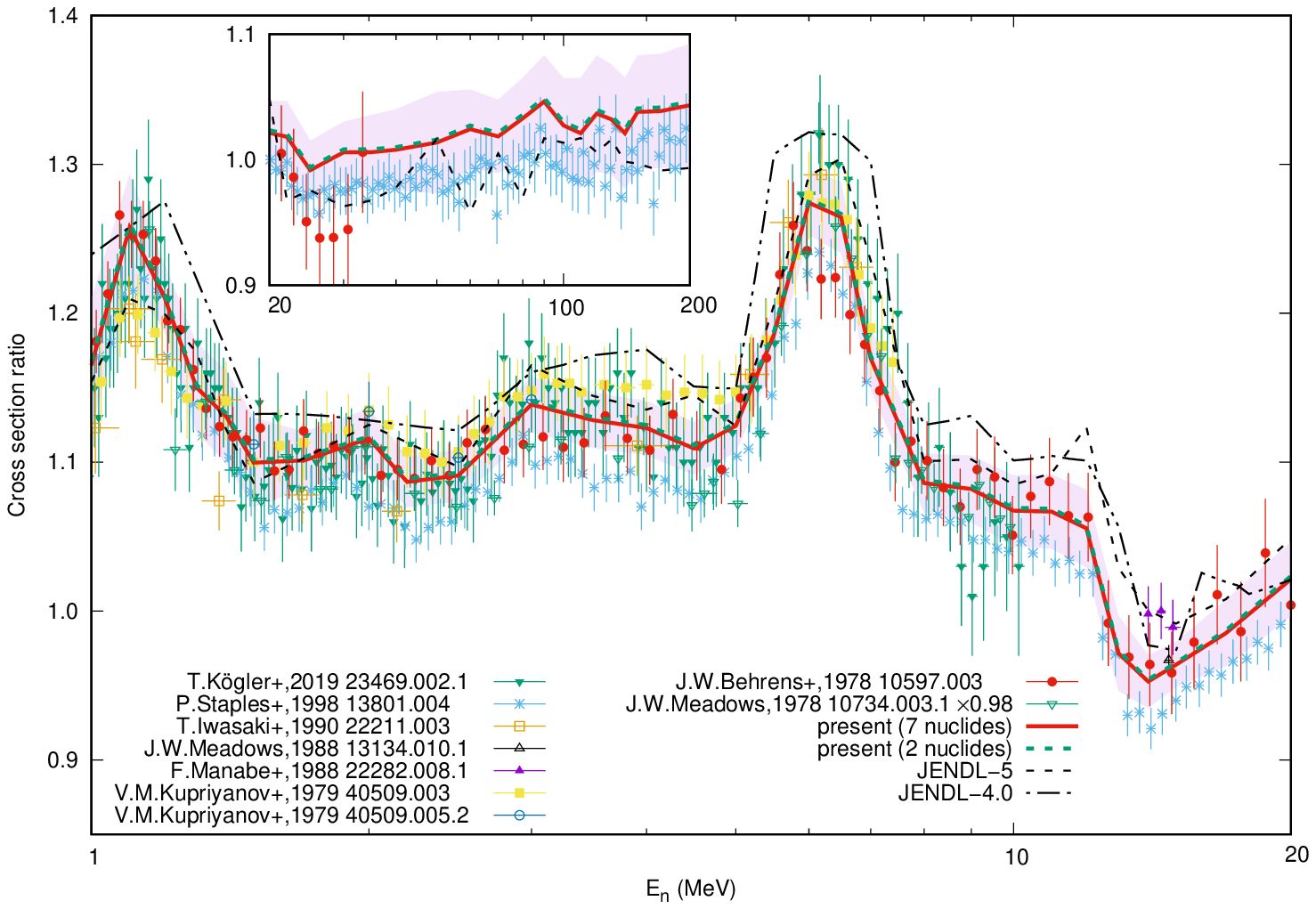}
\caption{$^{242}$Pu/$^{235}$U fission cross section ratios above 1~MeV from evaluations along with the experimental ones used in the present evaluation\cite{
Koegler2019Fast,
Staples1998Neutron,
Iwasaki1990Measurement,
Meadows1988Fission,
Manabe1988Measurements,
Kupriyanov1979Measurement,
Behrens1978Measurements,
Meadows1978Fission
}.
}
\label{fig:Pu242U235-hig}
\end{figure}

\begin{figure}
\centering\includegraphics[clip,angle=-90,width=\linewidth]{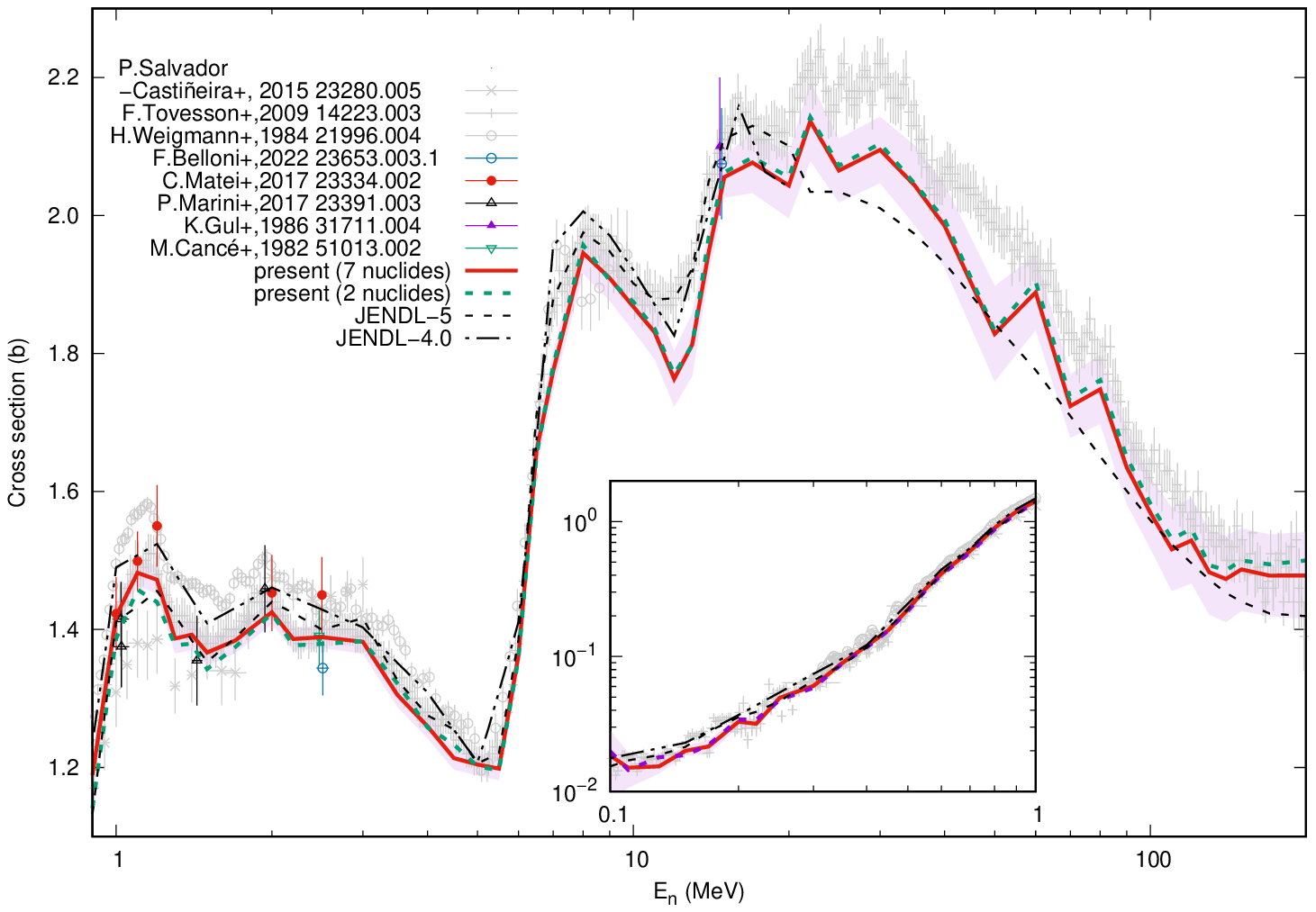}
\caption{$^{242}$Pu fission cross sections from evaluations along with the experimental ones used in the present evaluation\cite{
Belloni2022Neutron,
Matei2017Absolute,
Marini2017242Pu,
Gul1986Measurements,
Cance1983Mesures
}.
Three datasets excluded from the present evaluation~\cite{Salvador-Castineira2015bNeutron,Tovesson2009Neutron,Weigmann1985Neutron} are also plotted by grey symbols.
}
\label{fig:Pu242}
\end{figure}

\begin{figure}
\centering\includegraphics[clip,angle=-90,width=\linewidth]{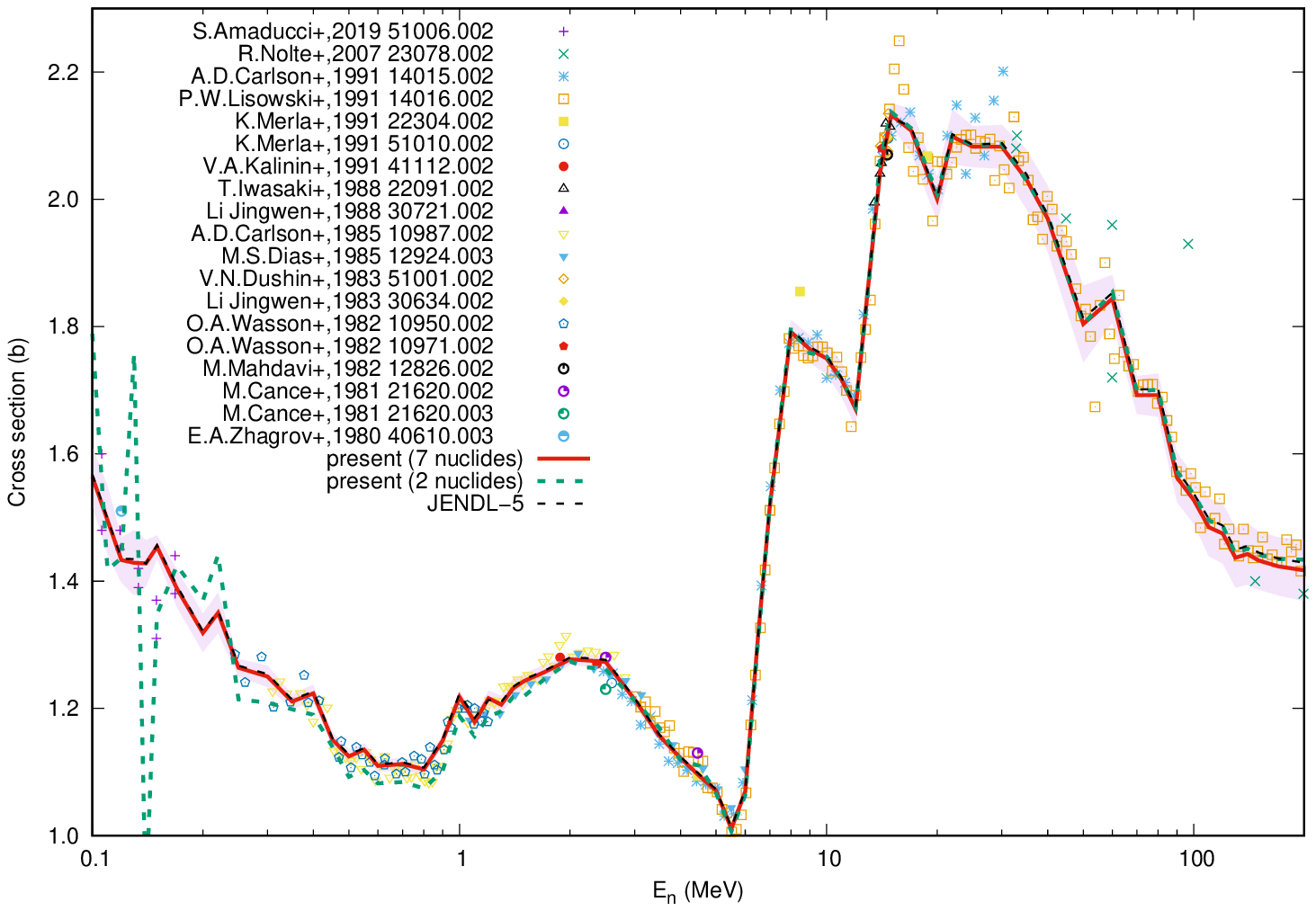}
\caption{$^{235}$U fission cross sections from evaluations along with the experimental ones used in the present evaluation\cite{
Amaducci2019Measurement,
Nolte2007Cross,
Carlson1992Measurements,
Lisowski1991Fission,
Merla1991Absolute,
Kalinin1991Correction,
Iwasaki1988Measurement,
Li1988Absolute,
Carlson1985Absolute,
Dias1985Application,
Dushin1983Statistical,
Li1983Absolute,
Wasson1982Absolute,
Wasson1982Measurement,
Mahdavi1983Measurements,
Cance1981Measures,
Zhagrov1980Fission
}.
Error bars of the experimental data points are omitted for readability.
}
\label{fig:U235}
\end{figure}

The newly evaluated $^{242}$Pu cross section and its ratio to the $^{235}$U cross section are plotted in Figs.~\ref{fig:Pu242U235-low} to \ref{fig:Pu242} along with the $^{235}$U cross section in Fig.~\ref{fig:U235}
\footnote{
The evaluated cross sections in an ASCII file are available as a Supplemental Material of this article.
}.
The symbols in these figures show the experimental data points used in the present evaluation except for the three $^{242}$Pu cross section datasets plotted in Fig.~\ref{fig:Pu242} in grey symbols.
The band accompanying the newly evaluated cross section or its ratio in each figure shows the external uncertainty (\ie uncertainty obtained by the least-squares analysis multiplied by the square root of the reduced chi-square) in the cross section from the seven-nuclide evaluation.

Figure~\ref{fig:Pu242U235-low} shows the $^{242}$Pu/$^{235}$U cross section ratio below 1~MeV.
Only two experimental datasets are available below 400~keV and they do not agree with each other within their error bars.
The ratio from the present evaluation is between these two experimental datasets,
while the JENDL-5 ratio is closer to Behrens \etal. rather than Kupriyanov \etal.
This figure shows that all evaluations and experimental datasets are consistent between 400~keV and 1~MeV.

Figure~\ref{fig:Pu242U235-hig} shows that the $^{242}$Pu/$^{235}$U ratio in JENDL-4.0 is systematically higher than the majority of the experimental datasets between 1 and 10~MeV.
Figure~\ref{fig:Pu242} in the same energy region shows that JENDL-4.0 is closer than the present evaluation to Weigmann's dataset.
We confirmed that the discrepancy between Weigmann's and other datasets is not resolved even if we renormalise Weigmann's dataset (originally normalised with the ENDF/B-V $^{235}$U fission cross section) with the latest IAEA standard~\cite{Carlson2018Evaluation}.
As the $^{242}$Pu/$^{235}$U ratio from Weigmann's measurement is not available in EXFOR,
this dataset is discarded in the present evaluation.
It is also not used in JENDL-5 evaluation since it is a dataset released before 2000.
Figure~\ref{fig:Pu242U235-hig} shows that the JENDL-5 ratio is systematically higher than the present ratio above 10~MeV though they agree with each other within the uncertainty associated with the present evaluation in general.

The systematic deviation between present evaluation and Staple \etal.'s measurement seen in Fig.~\ref{fig:Pu242U235-hig} is remarkable.
The inset of Fig.~\ref{fig:Pu242U235-hig} shows that the $^{242}$Pu/$^{235}$U ratio from the present evaluation above 20~MeV is systematically higher than the ratio measured by Staples \etal., which is more consistent with the JENDL-5 evaluation.
The main part of Fig.~\ref{fig:Pu242U235-hig} also shows that the present ratio is systematically higher than Staple \etal.'s ratio above 10~MeV.
The $^{242}$Pu/$^{235}$U ratio in this region was also measured by the CERN n\_TOF fission ionization chamber~\cite{Tsinganis2014aMeasurement,Tsinganis2014bMeasurement}.
The numerical data of the n\_TOF ratio have not been submitted to the EXFOR library yet and we could not use them for our evaluation.
A figure in Tsinganis's thesis~\cite{Tsinganis2014cMeasurement} shows the n\_TOF ratio is more consistent with the measurements by Behrens \etal. and Manabe \etal. rather than Staple \etal. between 10 and 20~MeV,
and hence we infer our present evaluation is also consistent with the n\_TOF ratio in this energy region.

Figure~\ref{fig:Pu242} shows that the present evaluation is consistent with the experimental absolute $^{242}$Pu cross sections adopted in the present evaluation including the three recent measurements by Belloni \etal., Matei \etal. and Marini \etal. though the cross section measured by Matei \etal. is higher than ours.
Between 1 and 2~MeV,
Figs.~\ref{fig:Pu242} and ~\ref{fig:U235} show that the two-nuclide evaluation makes the $^{242}$Pu and $^{235}$U cross sections lower than those from the seven-nuclide evaluation,
while Fig.~\ref{fig:Pu242U235-hig} shows that the evaluated $^{242}$Pu/$^{235}$U ratio does not show such difference between the two- and seven-nuclide evaluations.
It implies the evaluated $^{242}$Pu cross section is more strongly influenced by the experimental $^{242}$Pu/$^{235}$U cross section ratios than the experimental $^{242}$Pu absolute cross sections.

The $^{242}$Pu cross section measured by Tovesson \etal. is systematically higher than ours between 10 and 100~MeV as seen in Fig.~\ref{fig:Pu242}.
Tsinganis's thesis shows the n\_TOF $^{242}$Pu/$^{235}$U cross section ratio converted to the absolute $^{242}$Pu cross section is consistent with Tovesson \etal.'s cross section between 10 and 20~MeV where Tovesson \etal.'s cross section is more consistent with the JENDL-5 evaluation than the present evaluation.
This situation indicates that our evaluation could underestimate the actual fission cross section above 10~MeV
and we wish to repeat the evaluation again once the n\_TOF $^{242}$Pu/$^{235}$U ratio is compiled in EXFOR.

Figure~\ref{fig:U235} compares the $^{235}$U cross section updated by the present evaluation with JENDL-5 (our prior estimate).
This figure shows that the seven-nuclide evaluation does not introduce major change to the JENDL-5 $^{235}$U cross section below 100~MeV,
which is verified well with various important benchmark tests~\cite{Tada2023JENDL-5}.
On the other hand, the two-nuclide evaluation systematically reduces the prior estimate between 0.2 and 2~MeV.
The unphysical structure of the cross section between 0.1 and 0.2~MeV from the two-nuclide evaluation is resolved by inclusion of other five nuclides in evaluation.
Considering these facts,
we conclude that the seven-nuclide evaluation provides more reasonable evaluation of the $^{242}$Pu cross section.

\begin{figure}
\centering\includegraphics[clip,angle=-90,width=\linewidth]{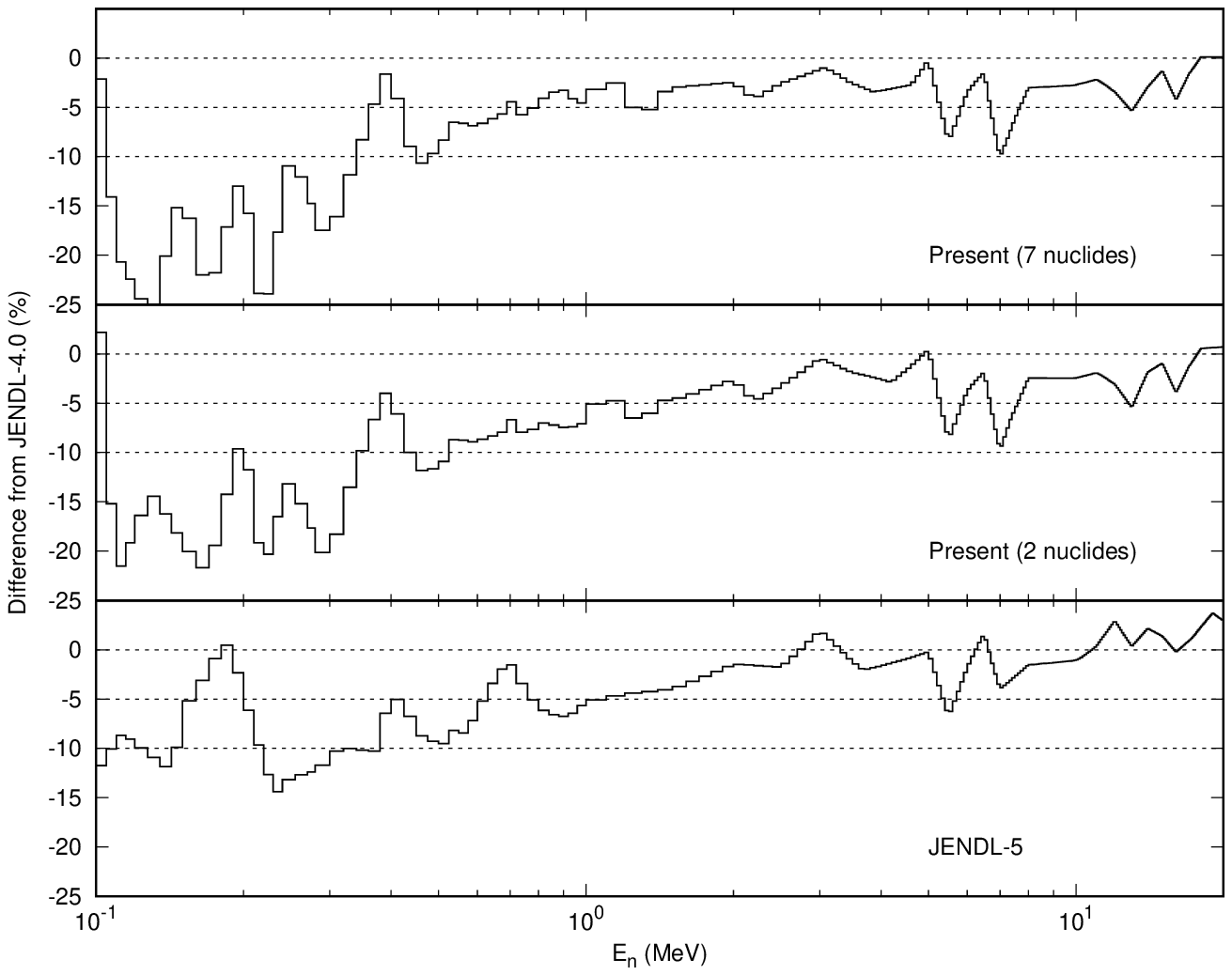}
\caption{Difference in $^{242}$Pu fission cross sections between evaluations.}
\label{fig:comp-Pu242-725g}
\end{figure}

Figure~\ref{fig:comp-Pu242-725g} shows the change in the new evaluation from the JENDL-4.0 evaluation in the SAND-II 725 energy group structure.
The change from the JENDL-4.0 evaluation looks similar for all three evaluations.
Below 2~MeV the decrease from the JENDL-4.0 evaluation becomes larger as the energy decreases 
and the difference is about 5\% around 1~MeV.

\subsection{Californium-252 spontaneous fission neutron spectrum averaged cross section}
\begin{figure}
\centering\includegraphics[clip,angle=-90,width=\linewidth]{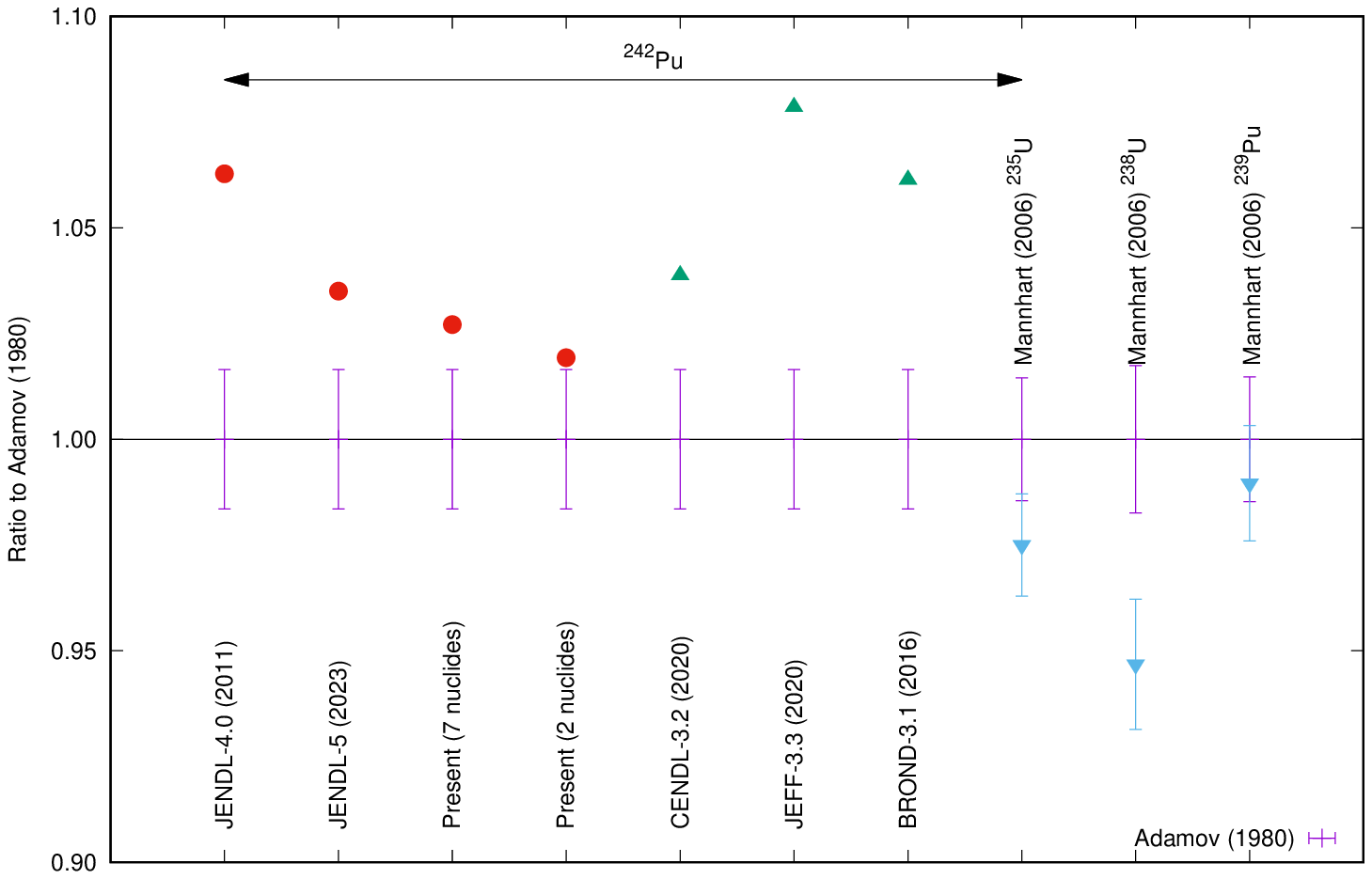}
\caption{Ratios of $^{252}$Cf spontaneous fission neutron spectrum averaged cross sections from present evaluation, evaluated data libraries~\cite{Shibata2011Japanese,Iwamoto2023Japanese,Ge2020CENDL,Plompen2020Joint,Blokhin2016New} and Mannhart's recommendation~\cite{Mannhart2006Response} to those measured by Adamov~\cite{Adamov1980Absolute}.
The ENDF/B-VIII.0~\cite{Brown2018ENDF} and TENDL-2021~\cite{Koning2019TENDL} libraries adopt the JENDL-4.0 evaluation and not shown. }
\label{fig:sacs-Cf252}
\end{figure}

Figure~\ref{fig:sacs-Cf252} shows the cross section averaged over the $^{252}$Cf spontaneous fission neutron spectrum (spectrum averaged cross section, SACS) relative to the SACS measured by Adamov~\cite{Adamov1980Absolute}. 
The evaluated cross sections from the SOK code were converted to the group-wise cross sections in the SAND-II 725 energy group structure and then averaged over the $^{252}$Cf spontaneous fission neutron spectrum in the 725 energy groups in the IRDFF-II library~\cite{Trkov2020IRDFF-II} as described in Ref.~\cite{Otuka2022aEXFOR}.
To obtain the SACS for the present results,
the newly evaluated cross section above 100~keV was merged with the cross section in the JENDL-5 library below 100~keV.

The JENDL-5 evaluation adopts the experimental datasets published after 2000 considering the trend that the newer measurements tend to report lower cross sections.
On the other hand,
the present simultaneous evaluation based on the measurements after 1970 gives lower SACS,
which are closer to the SACS measured by Adamov \etal. than the JENDL-5 evaluation.
The figure also shows that the SACS of CENDL-3.2~\cite{Ge2020CENDL} and JENDL-5 libraries are similar and those of other two libraries (JEFF-3.3~\cite{Plompen2020Joint} and BROND-3.1~\cite{Blokhin2016New}) are higher.
Note that the ENDF/B-VIII.0~\cite{Brown2018ENDF} and TENDL-2021~\cite{Koning2019TENDL} libraries adopt the JENDL-4.0 evaluation and they are not shown in this figure.

There is no other measured SACS in the EXFOR library for $^{242}$Pu fission.
In order to check reliability of Adamov \etal.'s measurement,
we also compared the widely accepted SACS recommended by Mannhart~\cite{Mannhart2006Response} with those measured by Adamov \etal. for the $^{235,238}$U and $^{239}$Pu fission cross sections.
We observe that Admov \etal.'s SACS is higher than Mannhart's recommendation in general.
Even if Adamov \etal.'s SACS are biased, however,
we can still conclude that the present evaluation shows better consistency with the measured and recommended SACS than the JENDL-4.0 library.

Table~\ref{tab:sacs-cf252} compares the SACS from various evaluations, measurements and recommendation.
This table shows that the good consistency between the JENDL-5 evaluation and Grundl \etal.'s measurement~\cite{Grundl1983Fission} (which is close to Mannhart's recommendation) is maintained by the present evaluation except for $^{238}$U cross section.
Our preceding evaluation for the JENDL-5 library~\cite{Otuka2022aEXFOR} and its minor update~\cite{Otuka2023Simultaneous} estimate the $^{238}$U SACS lower than those measured by Grundl and recommended by Mannhart,
and the present evaluation confirms that this underestimation still exists.
\begin{table}
\caption{$^{252}$Cf spontaneous fission neutron spectrum averaged cross sections (mb). See Ref.~\cite{Otuka2022bEXFOR} for the difference between Otuka (2022) and JENDL-5 (2023).}
\label{tab:sacs-cf252}
\begin{center}
\begin{tabular}{llllllll}
\hline
                                            &$^{242}$Pu  &$^{233}$U   &$^{235}$U   &$^{238}$U      &$^{239}$Pu  &$^{240}$Pu  &$^{241}$Pu  \\
\hline    
JENDL-4.0 (2011)~\cite{Shibata2011Japanese} & 1161       & 1906       & 1218       & 317           & 1802       & 1340       & 1625       \\           
JENDL-5 (2023)~\cite{Iwamoto2023Japanese}   & 1130       & 1901       & 1221       & 321           & 1808       & 1340       & 1606       \\
Otuka (2022)~\cite{Otuka2022aEXFOR}         &            & 1900       & 1223       & 316           & 1808       & 1340       & 1606       \\
Otuka (2023)~\cite{Otuka2023Simultaneous}   &            & 1899       & 1221       & 315           & 1814       & 1339       & 1608       \\
Present (7 nuclides)                        & 1122       & 1897       & 1220       & 315           & 1812       & 1337       & 1607       \\
Present (2 nuclides)                        & 1113       &            & 1206       &               &            &            &            \\
Adamov (1980)~\cite{Adamov1980Absolute}     & 1092$\pm$18& 1910$\pm$29& 1241$\pm$18& 344$\pm$6     & 1831$\pm$27&            &            \\
Mannhart (2006)~\cite{Mannhart2006Response} &            &            & 1210$\pm$15& 325.7$\pm$5.3 & 1812$\pm$25&            &            \\
Grundl (1983)~\cite{Grundl1983Fission}      &            & 1893$\pm$48& 1216$\pm$19& 326$ \pm$6.5  & 1824$\pm$35& 1337$\pm$32& 1616$\pm$80\\
\hline
\end{tabular}
\end{center}
\end{table}

\subsection{Verification with integral measurements}
\begin{figure}
\centering\includegraphics[clip,angle=-90,width=\linewidth]{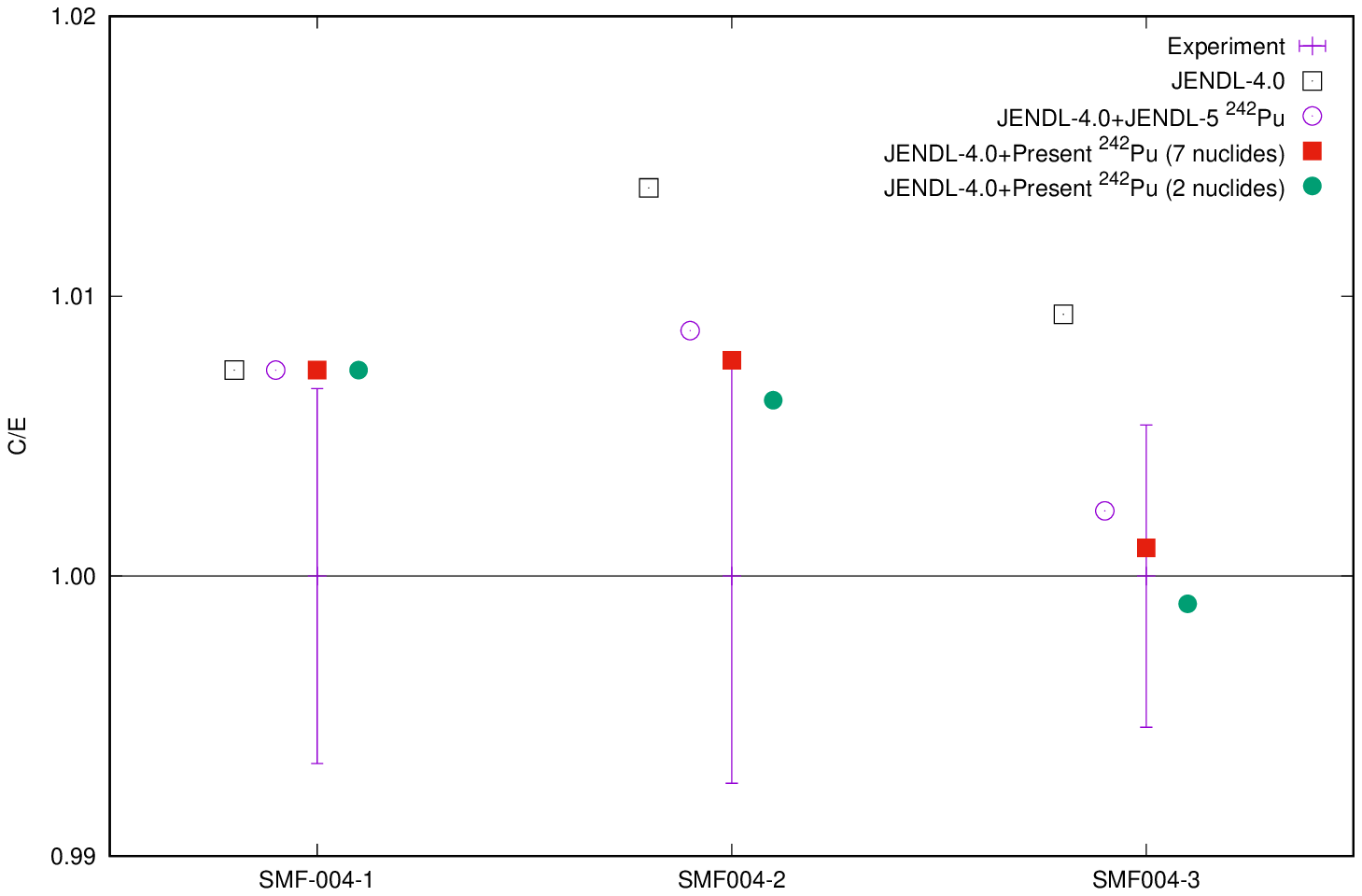}
\caption{C/E values of the LANL small-sized fast system (SPEC-MET-FAST-004-1, 2 and 3) criticalities calculated by ACE-FRENDY-CBZ sequence with the JENDL-4.0 library updated with the present $^{242}$Pu evaluation.
}
\label{fig:cbz-keff}
\end{figure}

To verify the newly evaluated $^{242}$Pu cross section, 
we calculated the criticalities ($k_\mathrm{eff}$) of three small-sized LANL fast systems known as SPEC-MET-FAST-004 Cases 1 to 3 in the ICSBEP Handbook.
\begin{itemize}
\item Case 1 ($^{239}$Pu): $^{239}$Pu plate without $^{242}$Pu.
\item Case 2 (Center-Driven): $^{239}$Pu plate sandwiched between high (94.9\%) and low (82.3\%) enriched $^{242}$Pu plates.
\item Case 3 (End-Driven): High and low enriched $^{242}$Pu plates sandwiched between $^{239}$Pu plates.
\end{itemize}
All these systems have cylinder shapes.
The criticalities were calculated by the ACE-FRENDY-CBZ neutronics analysis sequence~\cite{Chiba2023ACE} with the nuclear data libraries based on JENDL-4.0 except for $^{242}$Pu.
Energy-averaged (multi-group) macroscopic cross sections of each homogeneous region were calculated with FRENDY from the ACE-formatted continuous energy cross section data.
The VITAMIN-J 175-group structure was adopted and the neutron flux energy spectrum of the Godiva was used as the weight function.
The 175-group neutron transport equations for these systems were numerically solved by the discrete-ordinate solver SNRZ of CBZ.
The scattering anisotropy was taken into account by the Legendre polynomials up to the fifth order, and the S16 level symmetric angular quadrature set was used.
Figure~\ref{fig:cbz-keff} shows that the overestimation of the SMF004-2 and SMF004-3 criticalities with JENDL-4.0 is improved by adoption of the $^{242}$Pu fission cross section from the present work as well as the JENDL-5 library.

\section{Summary}
The $^{242}$Pu neutron-induced fission cross section was evaluated between 100~keV and 200~MeV.
The experimental $^{242}$Pu fission cross sections and $^{242}$Pu/$^{235}$U fission cross section ratios in the EXFOR library were reviewed and analysed simultaneously.
In addition to the evaluation including the two nuclides,
we also performed evaluation including the seven nuclides by extending our experimental database to the $^{233,238}$U and $^{239,240,241}$Pu fission cross sections and their ratios.
The newly obtained $^{242}$Pu fission cross section is systematically lower than the JENDL-4.0 cross section in the energy region of the prompt fission neutron spectrum peak region, and its reduction is about 5\% around 1~MeV.
The $^{242}$Pu fission cross sections from our evaluation and JENDL-5 evaluation are close to each other below 1~MeV while systematically differ from each other above 10~MeV.
The $^{252}$Cf fission neutron spectrum averaged cross section (SACS) of the $^{242}$Pu fission cross section from the present evaluation improves the overestimation of Adamov \etal.'s experimental SACS by the JENDL-4.0 cross section.
The newly obtained cross section also improves overestimation of the criticalities of $^{242}$Pu enriched systems by the JENDL-4.0 cross section on the same level with the JENDL-5 cross section.

\section*{Acknowledgement}
Antonio Jim\'{e}nez-Carrascosa and Oscar Cabellos (Universidad Polit\'{e}cnica de Madrid) performed criticality calculations by KENO to check if our criticality calculations are reasonable.
We thank Melissa Denecke (IAEA) for careful reading of the manuscript.
RO would like to thank the members of IAEA Nuclear Data Section for their hospitality during her internship.
Her internship was financially supported by ``Fundamental Nuclear Education Program by Japanese University Network for Global Nuclear Human Resource Development" entrusted to Tokyo Institute of Technology by Ministry of Education, Culture, Sports, Science and Technology (MEXT).
\bibliography{oku-jnst}

\begin{thebibliography}{10}
\providecommand{\url}[1]{\normalfont{#1}}
\providecommand{\urlprefix}{Available from: }

\bibitem{Salvatores2005Nuclear}
Salvatores~M. Nuclear fuel cycle strategies including {P}artitioning and
  {T}ransmutation. Nuclear Engineering and Design.
  2005;\hspace{0pt}235:805--816, DOI: 10.1016/j.nucengdes.2004.10.009.

\bibitem{Alberti2006Nuclear}
Aliberti~G, Palmiotti~G, Salvatores~M, et~al. Nuclear data sensitivity,
  uncertainty and target accuracy assessment for future nuclear systems. Annals
  of Nuclear Energy. 2006;\hspace{0pt}33:700--733, DOI:
  10.1016/j.anucene.2006.02.003.

\bibitem{Salvatores2008Uncertainty}
Salvatores~M, Aliberti~G, Dunn~M, et~al. Uncertainty and target accuracy
  assessment for innovative systems using recent covariance data evaluations.
  Nuclear Energy Agency; 2008.  NEA/WPEC-26.

\bibitem{Igashira2014Nuclear}
Igashira~M, Katabuchi~T, Harada~H, et~al. {A} nuclear data project on neutron
  capture cross sections of long-lived fission products and minor actinides.
  Nuclear Data Sheets. 2014;\hspace{0pt}118:72--77, DOI:
  10.1016/j.nds.2014.04.006.

\bibitem{Iwamoto2009JENDL}
Iwamoto~O, Nakagawa~T, Otuka~N, et~al. {JENDL Actinoid File 2008}. Journal of
  Nuclear Science and Technology. 2009;\hspace{0pt}46:510--528, DOI:
  10.1080/18811248.2007.9711557.

\bibitem{Mukaiyama1986Actinide}
Mukaiyama~T, Ohbu~M, Nakano~M, et~al. Actinide integral measurements in {FCA}
  assemblies. Radiation Effects. 1986;\hspace{0pt}93:147--152, DOI:
  10.1080/00337578608207441.

\bibitem{Chiba2011JENDL-4.0}
Chiba~G, Okumura~K, Sugino~K, et~al. {JENDL}-4.0 benchmarking for fission
  reactor applications. Journal of Nuclear Science and Technology.
  2011;\hspace{0pt}48:172--187, DOI: 10.1080/18811248.2011.9711692.

\bibitem{Tada2023JENDL-5}
Tada~K, Nagaya~Y, Taninaka~H, et~al. {JENDL-5} benchmarking for fission reactor
  applications. Journal of Nuclear Science and Technology.
  2023;\hspace{0pt}60:???--???, DOI: 10.1080/00223131.2023.2197439.

\bibitem{Martin2022Nuclear}
Martin~M, Nesaraja~C. Nuclear data sheets for {A}=242. Nuclear Data Sheets.
  2022;\hspace{0pt}186:263--395, DOI: 10.1016/j.nds.2022.11.003.

\bibitem{Mughabghab2018Atlas}
Mughabghab~S. {Atlas of Neutron Resonances} (sixth edition). Elsevier Science;
  2018.

\bibitem{Tovesson2009Neutron}
Tovesson~F, Hill~TS, Mocko~M, et~al. Neutron induced fission of
  $^{240,242}${Pu} from 1~{eV} to 200 {MeV}. Physical Review C.
  2009;\hspace{0pt}79:014613, DOI: 10.1103/PhysRevC.79.014613. EXFOR 14223.

\bibitem{Belloni2022Neutron}
Belloni~F, Eykens~R, Heyse~J, et~al. Neutron induced fission cross section
  measurements of $^{240}${Pu} and $^{242}${Pu} relative to the neutron--proton
  scattering cross section at 2.5 and 14.8~{MeV}. The European Physical Journal
  A. 2022;\hspace{0pt}58:227, DOI:10.1140/epja/s10050--022--00858--9. EXFOR
  23653.

\bibitem{Koegler2019Fast}
K\"{o}gler~T, Junghans~A, Beyer~R, et~al. Fast-neutron-induced fission cross
  section of $^{242}${Pu} measured at the neutron time-of-flight facility
  $n${ELBE}. Physical Review C. 2019;\hspace{0pt}99:024604, DOI:
  10.1103/PhysRevC.99.024604. EXFOR 23469.

\bibitem{Marini2017242Pu}
Marini~P, Mathieu~L, A\"{i}che~M, et~al. $^{242}${Pu} neutron-induced fission
  cross-section measurement from 1 to 2~{MeV} neutron energy. Physical Review
  C. 2017;\hspace{0pt}96:054604, DOI: 10.1103/PhysRevC.96.054604. EXFOR 23391.

\bibitem{Matei2017Absolute}
Matei~C, Belloni~F, Heyse~J, et~al. Absolute cross section measurements of
  neutron-induced fission of $^{242}${Pu} from 1 to 2.5~{MeV}. Physical Review
  C. 2017;\hspace{0pt}95:024606, DOI: 10.1103/PhysRevC.95.024606. EXFOR 23334.

\bibitem{Salvador-Castineira2015bNeutron}
Salvador-Casti{\~n}eira~P, Bry\'{s}~T, Eykens~R, et~al. Neutron-induced fission
  cross sections of $^{242}${Pu} from 0.3 {MeV} to 3~{MeV}. Physical Review C.
  2015;\hspace{0pt}92:044606, DOI: 10.1103/PhysRevC.92.044606. EXFOR 23280.

\bibitem{Shibata2011Japanese}
Shibata~K, Iwamoto~O, Nakagawa~T, et~al. {JENDL}-4.0: {A} new library for
  nuclear science and engineering. Journal of Nuclear Science and Technology.
  2011;\hspace{0pt}48:1--30, DOI: 10.1080/18811248.2011.9711675.

\bibitem{Fukushima2017Analyses}
Fukushima~M, Tsujimoto~K, Okajima~S. Analyses with latest major nuclear data
  libraries of the fission rate ratios for several {TRU} nuclides in the
  {FCA-IX} experiments. Journal of Nuclear Science and Technology.
  2017;\hspace{0pt}54:795--805, DOI: 10.1080/00223131.2017.1299644.

\bibitem{Iwamoto2023Japanese}
Iwamoto~O, Iwamoto~N, Kunieda~S, et~al. Japanese evaluated nuclear data library
  version 5: {JENDL-5}. Journal of Nuclear Science and Technology.
  2023;\hspace{0pt}60:1--60, DOI: 10.1080/00223131.2022.2141903.

\bibitem{Carlson2018Evaluation}
Carlson~A, Pronyaev~V, Capote~R, et~al. Evaluation of the neutron data
  standards. Nuclear Data Sheets. 2018;\hspace{0pt}148:143--188, DOI:
  10.1016/j.nds.2018.02.002.

\bibitem{Carlson2009International}
Carlson~A, Pronyaev~V, Smith~D, et~al. International evaluation of neutron
  cross section standards. Nuclear Data Sheets.
  2009;\hspace{0pt}110:3215--3324, DOI: 10.1016/j.nds.2009.11.001.

\bibitem{IAEA2007International}
{International Atomic Energy Agency}. International evaluation of neutron
  cross-section standards. International Atomic Energy Agency; 2007.
  STI/PUB/1291.

\bibitem{Otuka2022aEXFOR}
Otuka~N, Iwamoto~O. {EXFOR}-based simultaneous evaluation of neutron-induced
  uranium and plutonium fission cross sections for {JENDL-5}. Journal of
  Nuclear Science and Technology. 2022;\hspace{0pt}59:1004--1036, DOI:
  10.1080/00223131.2022.2030259.

\bibitem{Kawano2000Simultaneous}
Kawano~T, Matsunobu~H, Murata~T, et~al. Simultaneous evaluation of fission
  cross sections of uranium and plutonium isotopes for {JENDL}-3.3. Journal of
  Nuclear Science and Technology. 2000;\hspace{0pt}37:327--334, DOI:
  10.1080/18811248.2000.9714902.

\bibitem{Kawano2000Evaluation}
Kawano~T, Matsunobu~H, Murata~T, et~al. Evaluation of fission cross sections
  and covariances for $^{233}${U}, $^{235}${U}, $^{238}${U}, $^{239}${Pu},
  $^{240}${Pu}, and $^{241}${Pu}. Japan Atomic Energy Research Institute; 2000.
   JAERI-Research 2000-004. DOI: 10.11484/jaeri-research-2000-004.

\bibitem{Otuka2014Towards}
Otuka~N, Dupont~E, Semkova~V, et~al. Towards a more complete and accurate
  experimental nuclear reaction data library ({EXFOR}): International
  collaboration between {N}uclear {R}eaction {D}ata {C}entres ({NRDC}). Nuclear
  Data Sheets. 2014;\hspace{0pt}120:272--276, DOI: 10.1016/j.nds.2014.07.065.

\bibitem{Weigmann1985Neutron}
Weigmann~H, Wartena~J, B\"{u}rkholz~C. Neutron-induced fission cross section of
  $^{242}${Pu}. Nuclear Physics A. 1985;\hspace{0pt}438:333--353, DOI:
  10.1016/0375--9474(85)90379--3. EXFOR 21931, 21996.

\bibitem{Bergen1971Neutron}
Bergen~D, Fullwood~R. Neutron-induced fission cross section of $^{242}${Pu}.
  Nuclear Physics A. 1971;\hspace{0pt}163:577--582, DOI:
  10.1016/0375--9474(71)90510--0. EXFOR 10062.

\bibitem{Auchampaugh1971Neutron}
Auchampaugh~G, Farrell~J, Bergen~D. Neutron-induced fission cross sections of
  $^{242}${Pu} and $^{244}${Pu}. Nuclear Physics A.
  1971;\hspace{0pt}171:31--43, DOI: 10.1016/0375--9474(71)90360--5. EXFOR
  10266.

\bibitem{Khan198014.8}
Khan~N, Khan~H, Gul~K, et~al. 14.8~{MeV} neutron induced fission studies of
  $^{239}${Pu}, $^{242}${Pu}, $^{244}${Pu} and $^{241}${Am}. Nuclear
  Instruments and Methods. 1980;\hspace{0pt}173:163--168, DOI:
  10.1016/0029--554X(80)90583--2. EXFOR 30548.

\bibitem{Shpak1990Angular}
Shpak~D. Angular anisotropy of fragments from fission of $^{242}${Pu} by
  0.10--6.38~{MeV} neutrons. Soviet Journal of Nuclear Physics.
  1990;\hspace{0pt}52:419--423. EXFOR 41098.

\bibitem{Fursov1973Anomaly}
Fursov~B, Kuprianov~V, Smirenkin~G. Anomaly of the (n,n'f) reaction threshold.
  JETP Letters. 1973;\hspace{0pt}17:257--261. EXFOR 41656.

\bibitem{Otuka2022bEXFOR}
Otuka~N, Iwamoto~O. {EXFOR}-based simultaneous evaluation of neutron-induced
  uranium and plutonium fission cross sections for {JENDL-5}: {I}nputs and
  outputs. Japan Atomic Energy Agency; 2022.  JAEA-Data/Code 2022-005. DOI:
  10.11484/jaea-data-code-2022-005.

\bibitem{Otuka2023Simultaneous}
Otuka~N, Iwamoto~O. Simultaneous evaluation of uranium and plutonium fast
  neutron fission cross sections up to 200~{MeV} for {JENDL-5} and its updates.
  EPJ Web of Conferences. 2023;\hspace{0pt}284:08011, DOI:
  10.1051/epjconf/202328408011.

\bibitem{Tsinganis2014aMeasurement}
Tsinganis~A, Berthoumieux~E, Guerrero~C, et~al. Measurement of the
  $^{242}${Pu}(n,f) cross section at {n\_TOF}. EPJ Web of Conferences.
  2014;\hspace{0pt}66:03088, DOI: 10.1051/epjconf/20146603088.

\bibitem{Tsinganis2014bMeasurement}
Tsinganis~A, Berthoumieux~E, Guerrero~C, et~al. Measurement of the
  $^{242}${Pu}(n,f) cross section at the {CERN} {n\_TOF} facility. Nuclear Data
  Sheets. 2014;\hspace{0pt}119:58--60, DOI: 10.1016/j.nds.2014.08.018.

\bibitem{Tsinganis2014cMeasurement}
Tsinganis~A. Measurement of the $^{242}${Pu}(n,f) reaction cross-section at the
  {CERN} {n\_TOF} facility [dissertation]. submitted to National Technical
  University of Athens; 2014.

\bibitem{Adamov1980Absolute}
Adamov~V, Alkhazov~I, Gusev~S, et~al. Absolute measurements of induced fission
  cross sections of heavy nuclides for both $^{252}${Cf} fission spectrum
  neutrons and 14.7-{MeV} neutrons. In: Proceedings of the International
  Conference on Nuclear Cross Sections for Technology, Knoxville, 22--26
  October 1979; 1980. p. 995--999; NBS Special Publication 594. 40547.

\bibitem{Trkov2020IRDFF-II}
Trkov~A, Griffin~P, Simakov~S, et~al. {IRDFF-II}: A new neutron metrology
  library. Nuclear Data Sheets. 2020;\hspace{0pt}163:1--108, DOI:
  10.1016/j.nds.2019.12.001.

\bibitem{Ge2020CENDL}
Ge~Z, Xu~R, Wu~H, et~al. {CENDL-3.2}: The new version of {C}hinese general
  purpose evaluated nuclear data library. EPJ Web of Conferences.
  2020;\hspace{0pt}239:09001, DOI: 10.1051/epjconf/202023909001.

\bibitem{Plompen2020Joint}
Plompen~AJM, Cabellos~O, De~Saint~Jean~C, et~al. The joint evaluated fission
  and fusion nuclear data library, {JEFF-3.3}. The European Physical Journal A.
  2020;\hspace{0pt}56:181, DOI: 10.1140/epja/s10050--020--00141--9.

\bibitem{Blokhin2016New}
Blokhin~A, Gai~E, Ignatyuk~A, et~al. New version of neutron data library
  {BROND-3.1}. Voprosy Atomnoy Nauki i Tekhniki, Seriya Yaderno-Reaktornye
  Konstanty. 2016;\hspace{0pt}(2):62--93. In Russian.

\bibitem{Brown2018ENDF}
Brown~D, Chadwick~M, Capote~R, et~al. {ENDF/B-VIII.0}: The 8th major release of
  the nuclear reaction data library with {CIELO}-project cross sections, new
  standards and thermal scattering data. Nuclear Data Sheets.
  2018;\hspace{0pt}148:1--142, DOI: 10.1016/j.nds.2018.02.001.

\bibitem{Koning2019TENDL}
Koning~A, Rochman~D, Sublet~JC, et~al. {TENDL}: Complete nuclear data library
  for innovative nuclear science and technology. Nuclear Data Sheets.
  2019;\hspace{0pt}155:1--55, DOI: 10.1016/j.nds.2019.01.002.

\bibitem{Mannhart2006Response}
Mannhart~W. Response of activation reactions in the neutron field of
  californium-252 spontaneous fission. International Atomic Energy Agency;
  2006.  STI/DOC/010/452. {p}.~30--45.

\bibitem{Grundl1983Fission}
Grundl~J, Gilliam~D. Fission cross-section measurements in reactor physics and
  dosimetry benchmarks. Transactions of the American Nuclear Society.
  1983;\hspace{0pt}44:533--535. EXFOR 10809,12821.

\bibitem{Chiba2023ACE}
Chiba~G, Yamamoto~A, Tada~K. {ACE-FRENDY-CBZ:} a new neutronics analysis
  sequence using multi-group neutron transport calculations. Journal of Nuclear
  Science and Technology. 2023;\hspace{0pt}60:132--139, DOI:
  10.1080/00223131.2022.2087783.

\bibitem{Meadows1978Fission}
Meadows~J. The fission cross sections of plutonium-239 and plutonium-242
  relative to uranium-235 from 0.1 to 10~{MeV}. Nuclear Science and
  Engineering. 1978;\hspace{0pt}68:360--363, DOI: 10.13182/NSE78--A27315. EXFOR
  10734.

\bibitem{Schwerer2014EXFOR}
Schwerer~O, Otuka~N. {EXFOR/CINDA} dictionary manual. International Atomic
  Energy Agency; 2014.  IAEA-NDS-213 Rev. 2014/12.

\bibitem{Gul1986Measurements}
Gul~K, Ahmad~M, Anwar~M, et~al. Measurements of neutron fission cross sections
  of $^{237}${Np}, $^{240}${Pu}, $^{241}${Pu}, $^{242}${Pu}, $^{244}${Pu}, and
  $^{241}${Am} at 14.7~{MeV}. Nuclear Science and Engineering.
  1986;\hspace{0pt}94:42--45, DOI: 10.13182/NSE86--A17115. EXFOR 31711.

\bibitem{Cance1983Mesures}
Canc\'{e}~M, Grenier~G. Mesures absolues de $^{240}${Pu}(n,f),
  $^{242}${Pu}(n,f) et $^{237}${Np}(n,f) a l'energie incident de 2,5~{MeV}. In:
  Nuclear Data for Science and Technology, Proceedings of the International
  Conference, Antwerp 6--10 September 1982. Central Bureau for Nuclear
  Measurements; 1983. p. 51--54, DOI: 10.1007/978--94--009--7099--1\_10; EUR
  8355. In French. EXFOR 21821,51013.

\bibitem{Staples1998Neutron}
Staples~P, Morley~K. Neutron-induced fission cross-section ratios for
  $^{239}${Pu}, $^{240}${Pu}, $^{242}${Pu}, and $^{244}${Pu} relative to
  $^{235}${U} from 0.5 to 400~{MeV}. Nuclear Science and Engineering.
  1998;\hspace{0pt}129:149--163, DOI: 10.13182/NSE98--A1969. EXFOR 13801.

\bibitem{Iwasaki1990Measurement}
Iwasaki~T, Manabe~F, Baba~M, et~al. Measurement of fast neutron induced fission
  cross section ratios of {Pu}-240 and {Pu}-242 relative to {U}-235. Journal of
  Nuclear Science and Technology. 1990;\hspace{0pt}27:885--898, DOI:
  10.1080/18811248.1990.9731269. EXFOR 22211.

\bibitem{Meadows1988Fission}
Meadows~J. The fission cross sections of $^{230}${Th}, $^{232}${Th},
  $^{233}${U}, $^{234}${U}, $^{236}${U}, $^{238}${U}, $^{237}${Np},
  $^{239}${Pu} and $^{242}${Pu} relative to $^{235}${U} at 14.74~{MeV} neutron
  energy. Annals of Nuclear Energy. 1988;\hspace{0pt}15:421--429, DOI:
  10.1016/0306--4549(88)90038--2. EXFOR 13134.

\bibitem{Manabe1988Measurements}
Manabe~F, Kanda~K, Iwasaki~T, et~al. Measurements of neutron induced fission
  cross section ratios of $^{232}${Th}, $^{233}${U}, $^{234}${U}, $^{236}${U},
  $^{238}${U}, $^{237}${Np}, $^{242}${Pu} and $^{243}${Am} relative to
  $^{235}${U} around 14~{MeV}. The Technology Reports of T\={o}hoku University.
  1988;\hspace{0pt}52:97--126. EXFOR 22282.

\bibitem{Kupriyanov1979Measurement}
Kupriyanov~V, Fursov~B, Maslennikov~B, et~al. Measurement of the
  $^{240}${Pu}/$^{235}${U} and $^{242}${Pu}/$^{235}${U} fission cross-section
  ratios for 0.127--7.4-{MeV} neutrons. Soviet Atomic Energy.
  1979;\hspace{0pt}46:35--39, DOI: 10.1007/BF01119949. EXFOR 40509.

\bibitem{Behrens1978Measurements}
Behrens~J, Newbury~R, Magana~J. Measurements of the neutron-induced fission
  cross sections of $^{240}${Pu}, $^{242}${Pu}, and $^{244}${Pu} relative to
  $^{235}${U} from 0.1 to 30~{MeV}. Nuclear Science and Engineering.
  1978;\hspace{0pt}66:433--441, DOI: 10.13182/NSE78--A27227. EXFOR 10597.

\bibitem{Arlt1981Absolute}
Arlt~R, Josch~M, Musiol~G, et~al. Absolute fission cross section measurement on
  $^{235}${U} at 8.4~{MeV} neutron energy. In: Proceedings of the X-th
  International Symposium on Selected Topics of the Interaction of Fast
  Neutrons and Heavy Ions with Atomic Nuclei, Gaussig, 17--21 November 1980.
  International Atomic Energy Agency; 1981. p. 35--39; INDC(GDR)-19. EXFOR
  31833.

\bibitem{Duran2019High}
Duran~I, Paradela~C, Caama{\~n}o~M, et~al. High-resolution evaluation of the
  {U5}(n,f) cross section from 3~{keV} to 30~{keV}. EPJ Web of Conferences.
  2019;\hspace{0pt}211:02003, DOI: 10.1051/epjconf/201921102003. EXFOR 23294.

\bibitem{Ren2023Measurement}
Ren~Z, Yang~Y, Liu~R, et~al. Measurement of the $^{236,238}${U}(n,f) cross
  sections from the threshold to 200~{MeV} at {CSNS Back-n}. The European
  Physical Journal A. 2023;\hspace{0pt}59:5, DOI:
  10.1140/epja/s10050--022--00910--8. EXFOR 32886.

\bibitem{Snyder2021Measurement}
Snyder~L, Anastasiou~M, Bowden~N, et~al. Measurement of the
  $^{239}${Pu}(n,f)/$^{235}${U}(n,f) cross-section ratio with the {NIFFTE}
  fission {Time Projection Chamber}. Nuclear Data Sheets.
  2021;\hspace{0pt}178:1--40. EXFOR 14721.

\bibitem{Amaducci2019Measurement}
Amaducci~S, Cosentino~L, Barbagallo~M, et~al. Measurement of the
  $^{235}${U}(n,f) cross section relative to the $^{6}${Li}(n,t) and
  $^{10}${B}(n,$\alpha$) standards from thermal to 170~{keV} neutron energy
  range at {n\_TOF}. The European Physical Journal A. 2019;\hspace{0pt}55:120,
  DOI: 10.1140/epja/i2019--12802--7. EXFOR 23453,51006.

\bibitem{Nolte2007Cross}
Nolte~R, Allie~MS, Brooks~FD, et~al. Cross sections for neutron-induced fission
  of $^{235}${U}, $^{238}${U}, $^{209}${Bi}, and $^{\rm nat}${Pb} in the energy
  range from 33 to 200~{MeV} measured relative to n-p scattering. Nuclear
  Science and Engineering. 2007;\hspace{0pt}156:197--210, DOI:
  10.13182/NSE06--14. EXFOR 23078.

\bibitem{Carlson1992Measurements}
Carlson~A, Wasson~O, Lisowski~P, et~al. Measurements of the $^{235}${U} cross
  section in the 3 to 30~{MeV} neutron energy region. In: Proceedings of the
  International Conference on Nuclear Data for Science and Technology,
  J\"ulich, 13--17 May 1991; 1992. p. 518--520, DOI:
  10.1007/978--3--642--58113--7\_147. EXFOR 14015.

\bibitem{Lisowski1991Fission}
Lisowski~P, Gavron~A, Parker~W, et~al. Fission cross sections in the
  intermediate energy region. In: Proceedings of a Specialists' Meeting on
  Neutron Cross Section Standards for the Energy Region above 20~{MeV},
  Uppsala, 21--23 May 1991; 1976. p. 177--186; NEANDC-305. EXFOR 14016.

\bibitem{Merla1991Absolute}
Merla~K, Hausch~P, Herbach~C, et~al. Absolute measurement of neutron induced
  fission cross-sections of $^{235}${U}, $^{238}${U}, $^{237}${Np} and
  $^{239}${Pu}. In: Proceedings of the International Conference on Nuclear Data
  for Science and Technology, J\"ulich, 13--17 May 1991; 1992. p. 510--513,
  DOI: 10.1007/978--3--642--58113--7\_145. EXFOR 22304,51010.

\bibitem{Kalinin1991Correction}
Kalinin~VA, Kuz'min~VN, Solin~LM, et~al. Correction to the results of absolute
  measurements of the $^{235}${U} fission cross section with 1.9 and 2.4-{MeV}
  neutrons. Soviet Atomic Energy. 1991;\hspace{0pt}71:700--704, DOI:
  10.1007/BF01121671. EXFOR 41112.

\bibitem{Iwasaki1988Measurement}
Iwasaki~T, Karino~Y, Matsuyama~S, et~al. Measurement of $^{235}${U} fission
  cross section around 14~{MeV}. In: Proceedings of the International
  Conference on Nuclear Data for Science and Technology, Mito, 30 May--3 June
  1988; 1988. p. 87--90. EXFOR 22091.

\bibitem{Li1988Absolute}
Li~J, Shen~G, Ye~Z, et~al. Absolute measurement of $^{235}${U} fission cross
  section induced by 14.2~{MeV} neutrons. Chinese Journal of Nuclear Physics.
  1988;\hspace{0pt}10:237--243. In Chinese. EXFOR 30721.

\bibitem{Carlson1985Absolute}
Carlson~A, Behrens~J, Johnson~R, et~al. Absolute measurements of the
  $^{235}${U}(n,f) cross section for neutron energies from 0.3 to 3~{MeV}. In:
  Proceedings of an Advisory Group Meeting on Nuclear Standard Reference Data,
  Geel, 12--16 November 1984. International Atomic Energy Agency; 1985. p.
  162--166; IAEA-TECDOC-335. EXFOR 10987.

\bibitem{Dias1985Application}
Dias~M, Carlson~A, Johnson~R, et~al. Application of the dual thin scintillator
  neutron flux monitor in a $^{235}${U}(n,f) cross-section measurement. In:
  Proceedings of an Advisory Group Meeting on Nuclear Standard Reference Data,
  Geel, 12--16 November 1984. International Atomic Energy Agency; 1985. p.
  467--470; IAEA-TECDOC-335. EXFOR 12924.

\bibitem{Dushin1983Statistical}
Dushin~VN, Fomichev~AV, Kovalenko~SS, et~al. Statistical analysis of
  experimental data on the cross sections of
  $^{233,235,238}${U},$^{237}${Np},$^{239,242}${Pu} fission by neutrons of
  energy 2.6, 8.5, and 14.5~{MeV}. Soviet Atomic Energy.
  1983;\hspace{0pt}55:656--660, DOI: 10.1007/BF01124127. EXFOR 51001.

\bibitem{Li1983Absolute}
Li~J, Li~A, Rong~C, et~al. Absolute measurements of fission cross sections for
  $^{235}${U} and $^{239}${Pu} induced by 14.7~{MeV} neutron using the
  associated particle method. Chinese Journal of Nuclear Physics.
  1983;\hspace{0pt}5:45--50. In Chinese. EXFOR 30634.

\bibitem{Wasson1982Absolute}
Wasson~O, Meier~M, Duvall~K. Absolute measurement of the uranium-235 fission
  cross section from 0.2 to 1.2~{MeV}. Nuclear Science and Engineering.
  1982;\hspace{0pt}81:196--212, DOI: 10.13182/NSE82--A20085. EXFOR 10950.

\bibitem{Wasson1982Measurement}
Wasson~O, Carlson~A, Duvall~K. Measurement of the $^{235}${U} neutron-induced
  fission cross section at 14.1~{MeV}. Nuclear Science and Engineering.
  1982;\hspace{0pt}80:282--303, DOI: 10.13182/NSE82--A21431. EXFOR 10971.

\bibitem{Mahdavi1983Measurements}
Mahdavi~M, Knoll~G, Robertson~J. Measurements of the 14~{MeV} fission
  cross-sections for $^{235}${U} and $^{239}${Pu}. In: Nuclear Data for Science
  and Technology, Proceedings of the International Conference, Antwerp 6--10
  September 1982. Central Bureau for Nuclear Measurements; 1983. p. 58--61,
  DOI: 10.1007/978--94--009--7099--1\_12; EUR 8355. EXFOR 12826.

\bibitem{Cance1981Measures}
Cance~M, Grenier~G. Mesures absolues des sections efficaces de fission de
  $^{235}${U} a 2,5~{MeV} et 4,5~{MeV} et de $^{241}${Am} a 14,6~{MeV}. Centre
  d'Etudes de Bruy\'{e}res-Ch\^{a}tel; 1981.  CEA-N-2194. In French. EXFOR
  21620.

\bibitem{Zhagrov1980Fission}
Zhagrov~E, Nemilov~Y, Platonov~A, et~al. Fission cross sections of $^{233}${U}
  and $^{235}${U} in intermediate neutron energy range. In: Proceedings of the
  5th All Union Conference on Neutron Physics, Kyiv, 15--19 September 1980;
  Vol.~3; 1980. p. 45--48. In Russian. EXFOR 40610.

\end{thebibliography}
\end{document}